\documentclass[pre,aps,reprint,superscriptaddress,floatflix,twocolumn,amsfonts,amsmath]{revtex4-1}
\usepackage{graphicx}
\usepackage{subfigure}
\usepackage{amssymb}
\usepackage{latexsym}
\usepackage{soul}
\usepackage{color}

\def\var{\mathop{\mathrm var}}

\begin{document}
\title{Stress-structure relation in dense colloidal melt under forward and instantaneous reversal of shear} 
\date{\today}
\def\unikn{\affiliation{Fachbereich Physik,
  Universit\"at Konstanz, 78457 Konstanz, Germany}}
\def\zk{\affiliation{Zukunftskolleg,
  Universit\"at Konstanz, 78457 Konstanz, Germany}}
\def\dlr{\affiliation{Institut f\"ur Materialphysik im Weltraum,
  Deutsches Zentrum f\"ur Luft- und Raumfahrt (DLR), 51170 K\"oln, Germany}}
\def\unidus{\affiliation{Institut f\"ur Theoretische Physik II,
  Soft Matter, Heinrich-Heine-Universit\"at D\"usseldorf,
  40225 D\"usseldorf, Germany}}
\def\courant{\affiliation{Courant Institute of Mathematical Sciences, 
  New York University, New York, NY 10012, USA}}

\author{Amit Kumar Bhattacharjee} 
\thanks{Present Address: Courant Institute of Mathematical Sciences, 
  New York University, New York, NY 10012}
\email{Email address: amitb@courant.nyu.edu}\unikn
%\author{Thomas Voigtmann}\dlr\zk

\begin{abstract}
Dense supercooled colloidal melt in forward shear from a quiescent state shows 
overshoot in shear stress at $10\%$ strain with an unchanged fluid structure at 
equal stress before and after overshoot. In addition, we find overshoot in normal 
stress with a monotonic increase in osmotic pressure at an identical strain. The 
first and second normal stress become comparable in magnitude and opposite in 
sign. Functional dependence of the steady state stress and osmotic pressure with 
P{\'e}clet number demonstrate signature of crossover between Newtonian and nearly-
Newtonian regime. Moreover, instantaneous shear reversal from steady state exhibit 
Bauschinger effect, where strong history dependence is observed depending on the 
time of flow reversal. The distribution of particulate stress and osmotic pressure 
at the point of flow reversal is shown to be a signature of the subsequent response. 
We link the history dependence of the stress-strain curves to changes in the fluid 
structure measured through the angular components of the radial distribution function. 
A uniform compression in transition from forward to reversed flowing state is found.
\end{abstract}

%\pacs{83.50.Ax, 83.60.Rs, 64.70.pv, 83.10.Mj}
\maketitle

\section{Introduction}
\label{sec:1}
The mechanical and structural response of dense colloidal dispersion under external 
forces are intriguing due to the interplay of associated single particle relaxation 
time scale $\tau_0$ and the structural relaxation time scale $\tau$, many orders of 
magnitude slower than the previous. A rich behaviour of amorphous solid like and 
non-Newtonian fluid like behaviour is seen in experiments and simulations when 
perturbed on a timescale $1/\dot\gamma$ with $\tau_0\ll\dot\gamma^{-1}\ll\tau$ \cite{biko}. 
Fluid like behaviour is exemplified by an increase in the shear viscosity 
referred as shear thinning or thickening, while the solid like behaviour through 
the finite yield stress and nonzero elastic constants of the media \cite{larson}.
Non-Newtonian fluids also exhibit nonzero normal stress components, signs of them 
lead to compression or propulsion of fluid mass leading to fascinating phenomena 
{\it e.g.} rotating rod flows \cite{BoPoGuazz} and rod climbing effects in attractive 
colloids \cite{FarReBrader}. The flowing steady state response is usually characterized 
as a nonlinear function of $\dot\gamma$ by the macroscopic shear stress $\sigma_{xy}$.
In addition, first and second normal stress together with osmotic pressure as a 
function of $\dot\gamma$ must be sought to characterize the stress tensor.
Nonlinear functional form of shear and normal stresses with P{\'e}clet number as 
well as a positive first normal stress and a sign reversed second normal stress 
for monodisperse colloidal suspensions at lower and moderately high volume fractions 
had been reported in Stokesian dynamics simulation of concentrated colloidal 
suspension \cite{phbrbo} as well as molecular dynamics (MD) simulation of 
non-Brownian spheres \cite{semochclbon}. Similar results are reported in system 
with charge-stabilized dispersion \cite{rawalu}, however the functional dependence 
in vicinity of glass transition are not known.

In addition to the study of steady state properties, rigorous theoretical, experimental 
as well as simulational emphasis has been employed in past couple of decades to study 
the transient response of various systems in external shear to understand the kinetic 
pathway through which these systems evolve to steady flowing state. As a constant 
strain rate is applied to a quiescent state at $t=0$, shear stress increase from 
zero to a steady-state value with an intermediate hump at $10\%$ strain amplitude 
for dense colloids, known in the literature as stress overshoot. These overshoot 
phenomena is reported in the MD simulations and experiments of gels \cite{kobabespobrpetek}, 
dense polymeric melts \cite{mocucocummin}, liquid crystalline polymers \cite{taotbrie} 
and charged as well as uncharged dense colloidal melt \cite{zaholaegbrvofu,frbhhofuvo,koschpetek}. 
The last example is reasonably well understood within the mode coupling theory (MCT) 
framework. An internal connection between superdiffusive particulate motion and jump 
in the local stress variances has been attributed to the stress overshoot. 

Considerable emphasis has also been exercised to understand a connection between 
shear-deformed microstructure and the macroscopic stresses \cite{kirk,HanRaHess,zaho}. 
A universal flow-induced structure at equal stresses have emerged at the elastic and 
plastic branch of the stress-strain curve. Application of shear deformation at constant 
rate leads to a modification of the stress overshoot in a system quenched into its 
glassy state: after startup flow, shear is reversed in the steady state leading to 
a vanishing of the maximum in the stress-strain curve known as the Bauschinger 
effect \cite{kalepr1,frbhhofuvo}. The effect has been analyzed in terms of anisotropic 
athermal elastic constants that arise since the initially isotropic amorphous state 
acquires anisotropy under the initial deformation. A gradual disappearance of the shear 
stress overshoot as well as ceasing superdiffusion is found by the successive flow 
reversal from intermediate states to the steady flowing state. The transient and steady 
state properties of osmotic pressure is difficult to measure in experiments and recently 
been computed in computer simulations \cite{BraLAOS}. To our knowledge, the transient 
and steady state response in terms of the macroscopic quantities and their relation 
to the shear-deformed microstructure under various flow history is yet to be sought.  

Here we investigate the nonlinear rheology under start up flow from equilibrated 
state as well as in instantaneous flow reversal from intermediate and steady 
flowing states, without referring to the athermal variant. The issues that we address 
here can be categorized into: (a) the transient and steady state response of the stresses 
and osmotic pressure to steady shear and response to instantaneous reversal of shear, 
(b) connection between Bauschinger effect and the particulate stresses and (c) the 
flow-induced structural response of the melt in transient and steady flowing states
and their relation with the macroscopic stresses. 

The article is organized as follows: Section~\ref{sec:2} gives a concise overview of 
the simulation method while Sec.~\ref{sec:3} pre-empts the results through an establishment 
of the connection between the mechanical response to the fluid microstructure. We discuss 
the central results in Sec ~\ref{sec:4} and Sec.~\ref{sec:5} concludes.

\section{Simulation Methods}
\label{sec:2}

We simulate a model system of dense supercooled colloidal suspension through nonequilibrium 
molecular dynamics simulation. An additive binary mixture of size ratio $1.2$ is chosen 
to avoid crystallization as well as artifacts due to large size disparity. The components 
interact within a cutoff distance $r < r_{c,\alpha\beta}(=2^{1/6}\sigma_{\alpha\beta})$ 
through a purely-repulsive soft-sphere WCA potential \cite{wechan}, $V_{\alpha\beta}(r)= 
4\epsilon_{\alpha\beta}\left[\left(\frac{\sigma_{\alpha\beta}}{r}\right)^{12}-\left(\frac{\sigma_{\alpha\beta}}{r}
\right)^6+\frac14\right]S_{\alpha\beta}(r)$ where $\sigma$ denotes the particle diameter 
and $\alpha,\beta$ the particle species. $S_{\alpha\beta}(r) =(r-r_{c,\alpha\beta})^4/[h^4+(r-r_{c,\alpha\beta})^4]$ 
with $h=10^{-2}\sigma_{\alpha\beta}$ is a smoothing function applied to ensure continuity 
of force and conservation of total energy in the NVE ensemble. Both species are assigned 
equal masses for convenience. Units of energy $\epsilon_{\alpha\beta}=1$ and units of 
length are so chosen that $\sigma_{\text{AA}}=1$ and the unit of time is $\tau=\sqrt{m_
\text{A}\sigma^2_{\text{AA}}/\epsilon_{\text{AA}}}$ where $m_\text{A}$ is the mass 
of A-species of particles. The simulation consisted of $N=2N_{\text{A}}=1300$ 
particles in a three-dimensional box with linear dimension $L=10\sigma_{\text{AA}}$, 
corresponding to a number density $\rho=1.3/\sigma_{\text{AA}}^3$ \cite{frbhhofuvo}.
 
Colloidal property is incorporated in the mixture by coupling to a dissipative particle 
dynamics (DPD) thermostat where the dissipative force, proportional to the relative 
velocity of two species of particles, ensures Galilean invariance thereby local 
conservation of momentum. The random force satisfying fluctuation dissipation relation 
ensures Boltzmann distribution in equilibrium. These two competing force sets uniquely 
the temperature of the system. The cut off radius for the thermostat is chosen to be 
$1.7r_{c,\alpha\alpha}$ and the controlling parameter for frictional forces is set to 
$\zeta=10$.  

The Langevin equation of motion are integrated with a generalized velocity Verlet 
algorithm with a time increment of $\delta t=5\times10^{-4}$. Our estimation of the 
glass transition point according to MCT description is $T_c\approx0.347$ and we focus 
our work on the equilibrated fluid at $T=0.4$. Initial equilibration is proceeded by 
step increments of $\delta t=10^{-3}$, assigning new velocities on every $50$ integration 
time steps. The simulation runs were long enough to observe the decay of the incoherent 
intermediate scattering function to zero for a wave number corresponding to a typical 
interparticle separation. A set of $1000$ independently equilibrated configurations 
served as initial configurations for the forward shear runs, while $1000$ pre-sheared 
configurations for three separate strains are chosen for shear reversal runs employing 
the DPD thermostat.

Shear is applied from strain-free configurations in the $x$-direction with a gradient 
in the $y$-direction (thus, vorticity along {\it z}-direction) at a fixed strain rate 
$\dot\gamma=5\times10^{-3}$ initially for $3\times10^7$ steps resulting to steady state 
with strain $\gamma=75$. Shear is reversed and proceeded from three pre-sheared states, 
denoted with strain $-\gamma_w^{el}=3.5\times10^{-2}$, $-\gamma_w^{max}=8.6\times10^{-2}$ 
and $-\gamma_w^{s}=75$ for $3\times10^7$ simulation steps, resulting to steady states 
with accumulated strain $-74.965, -74.914$ and $0$. For all these cases, planar Couette 
flow is imposed by periodic Lees-Edwards boundaries and an establishment of linear 
velocity profile is achieved within a few iterations. 

\section{Connection between stresses and microstructure}
\label{sec:3}

The amount of stress developed by the application of steady shear to a quiescent unsheared 
state or pre-sheared states are measured through the nonzero components of the stress 
tensor $\sigma_{\mu\nu}$. Kirkwood formula defines a combination of kinematic and virial 
contribution to the stress tensor \cite{kirk},
\begin{equation}
\label{eq.sstress}
  \sigma_{\mu\nu}=\langle\hat\sigma_{\mu\nu}\rangle
  =-\frac1V\left\langle\sum_{i=1}^N\left[m_i v_{i,\mu}v_{i,\nu}
  +\sum_{j\neq i}r_{ij,\mu}F_{ij,\nu}\right]\right\rangle\,
\end{equation}
where the angular brackets are indicative of canonical averaging, $i,j$ denote the particle
index and $\mu,\nu$ are the Cartesian directions. In a specified coordinate system having 
shear along {\it x} direction with a gradient along {\it y} direction, the dominant contribution 
to the shear stress is the off-diagonal $xy$-element and the normal stresses are the combination 
of the diagonal $xx, yy$ and $zz$-element of the stress tensor. The first and second normal 
stresses are the differences in the diagonal components as \cite{larson}, 
\begin{equation}
\label{eq.nstress}
\mathcal{N}_1 = \langle\sigma_{xx}-\sigma_{yy}\rangle; \; 
\mathcal{N}_2 = \langle\sigma_{yy}-\sigma_{zz}\rangle.
\end{equation}
and the third normal stress $\mathcal{N}_3$ is the sum of the two. The osmotic pressure is 
the sum of the diagonal elements of the stress tensor, 
\begin{equation} 
\label{eq:osmotic}
\mathcal{P} = -\frac{1}{3}\left\langle\sigma_{xx}+\sigma_{yy}+\sigma_{zz}\right\rangle.
\end{equation}  

The virial part of the stress tensor in Eq.(\ref{eq.sstress}) can be expressed in terms 
of the pair distribution function 
$g^{\alpha\beta}({\bf r})={V}/{N_\alpha N_\beta} \times \langle \sum_{i=1}^{N_\alpha} \sum_{j(\neq i)}^{N_\beta} \delta({\bf r}-|{\bf r}_i^\alpha - {\bf r}_j^\beta|) \rangle$
by rewriting the equation and substituting expressions for pairwise force in the following manner \cite{kirk}, 
\begin{eqnarray}
{\bf \sigma} &=& -\frac{1}{2V} \langle \sum_{i=1}^{N_\alpha} \sum_{j(\neq i)}^{N_\beta} {\bf r}_{ij} {\bf F}_{ij} \rangle, \\ \nonumber
             &=& -\frac{1}{2V} \langle \int d{\bf r} \sum_{\alpha,\beta} \sum_{i=1}^{N_\alpha} \sum_{j(\neq i)}^{N_\beta} \delta({\bf r}-|{\bf r}_i^\alpha - {\bf r}_j^\beta|) {\bf r} {\bf F}_{ij}\rangle, \\ \nonumber  
             &=& \frac{1}{2V} \langle \int d{\bf r} \sum_{\alpha,\beta} \sum_{i=1}^{N_\alpha} \sum_{j(\neq i)}^{N_\beta} \delta({\bf r}-|{\bf r}_i^\alpha - {\bf r}_j^\beta|) 
                 \frac{{\bf r}{\bf r}}{r} \frac{{\partial V}_{\alpha\beta}}{\partial r}\rangle, \\ \nonumber 
             &=& \frac{1}{2V^2} \int d{\bf r} \sum_{\alpha,\beta} N_\alpha N_\beta \frac{{\bf r}{\bf r}}{r} \frac{{\partial V}_{\alpha\beta}}{\partial r} g^{\alpha\beta}({\bf r}) \\ \nonumber
             &=& \frac{\rho^2}{2} \int d{\bf r} \sum_{\alpha,\beta} \frac{N_\alpha N_\beta}{N^2} \frac{{\bf r}{\bf r}}{r} \frac{{\partial V}_{\alpha\beta}}{\partial r} g^{\alpha\beta}({\bf r}) 
\end{eqnarray}
where $\rho = N/V$ is the average density and $N_{\alpha}, N_{\beta}$ correspond to the number of particles of the respective species index.

Shear induced pair distribution function for either forward or backward directed sheared 
states does not exhibit any significant structural change compared to the quiescent state. 
A more relevant quantity sensitive to shear is constructed \cite{kirk,HanRaHess,zaho}, 
where the three dimensional pair correlation function is expanded into the basis of spherical 
harmonics as, 
$g^{\alpha\beta}({\bf r})=\sum_l\sum_m g_{lm}^{\alpha\beta}(r)Y_{lm}(\theta,\phi)$. Here 
$Y_{lm}(\theta,\phi) = (-1)^m\sqrt{\frac{2l+1}{4\pi}\frac{(l-m)!}{(l+m)!}} P_{lm}(cos\theta) e^{im\phi}$ 
are the spherical harmonics of degree $l$ and order $m$, $\theta$ and $\phi$ are the polar and 
azimuthal angle and $P_{lm}(cos\theta)$ are the associated Legendre polynomial \cite{abrsteg}. 
From symmetry considerations, only even numbers in l is existent and the most relevant terms in 
the expansion are associated to $l=2,m=0,\pm2$. %which are given as $Y_{20}(\theta,\phi)=\sqrt{{5}/{16\pi}}{(2z^2-x^2-y^2)/{r^2}}$ and  
%$Y_{2\pm2}(\theta,\phi)=\sqrt{{15}/{32\pi}}{(x\pm iy)^2}/{r^2}$ respectively. 
Here we look for in-plain structural changes to shear and the relevant expansion coefficients 
associated to $g{(r)}$ are,
\begin{eqnarray}
\label{eq:Reg22}
Re[g_{22}^{\alpha\beta}(r)] 
&=& \sqrt{\frac{15}{16\pi}}\frac{V}{N_\alpha N_\beta}\Big\langle \sum_{i=1}^{N_\alpha} 
\sum_{j(\neq i)}^{N_\beta} \delta(r-|{\bf r}_i^\alpha - {\bf r}_j^\beta|) \nonumber \\ 
&& \times \frac{(x_i^\alpha-x_j^\beta)^2 - (y_i^\alpha-y_j^\beta)^2}{r^4} \Big\rangle \\
\label{eq:Img22}
Im[g_{22}^{\alpha\beta}(r)] 
&=& \sqrt{\frac{15}{8\pi}}\frac{V}{N_\alpha N_\beta}\Big\langle \sum_{i=1}^{N_\alpha} 
\sum_{j(\neq i)}^{N_\beta} \delta(r-|{\bf r}_i^\alpha - {\bf r}_j^\beta|) \nonumber \\ 
&& \times \frac{(x_i^\alpha-x_j^\beta)(y_i^\alpha-y_j^\beta)}{r^4} \Big\rangle
\end{eqnarray}
where $Re$ and $Im$ represent the real and imaginary part. Both these components are nonexistent 
in the quiescent state while they accumulate values in the sheared states. Integrating 
the algebraic combination of these functions together with the interparticle force yields the 
first normal stress \cite{ganeu} as well as the shear stress \cite{ganeu,zaho}, 
\begin{eqnarray}
\label{eq:N1}
\mathcal{N}_{1} &=& \rho^2\sqrt{\frac{2\pi}{15}}\sum_{\alpha\beta}\frac{N_\alpha N_\beta}{N^2} 
\int_0^\infty dr r^3 \frac{\partial V_{\alpha\beta}(r)}{\partial r} \; \nonumber \\ 
&& \times Re[g_{22}^{\alpha\beta}(r)],\\
\label{eq:sigma}
\sigma_{xy} &=& -\rho^2\sqrt{\frac{2\pi}{15}}\sum_{\alpha\beta}\frac{N_\alpha N_\beta}{N^2} 
\int_0^\infty dr r^3 \frac{\partial V_{\alpha\beta}(r)}{\partial r} \;  \nonumber \\ 
&& \times Im[g_{22}^{\alpha\beta}(r)].
\end{eqnarray}

\section{Results}
\label{sec:4}

We summarize the main findings in the dynamics and microstructure of the colloidal melt under 
shear startup and instantaneous shear reversal.

\subsection{Dynamics}
\subsubsection{Flow curve}
\begin{figure}
\centerline{\includegraphics[width=1.0\linewidth]{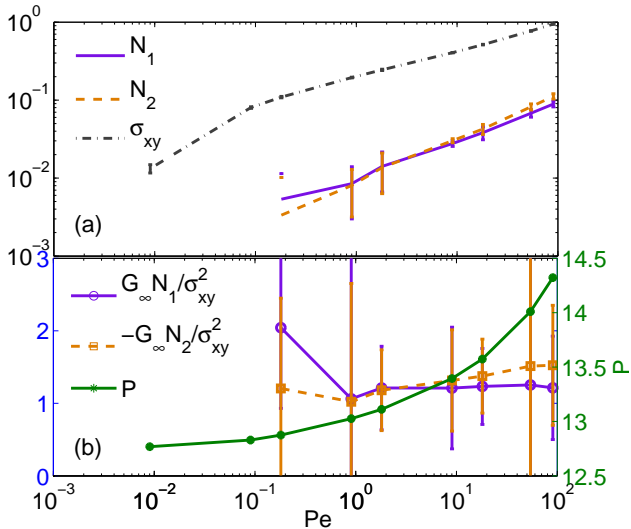}}
\caption{\label{fig:scaling}(Color online) Scaling of the steady state normal stresses 
  $\mathcal{N}_{1,2}$, shear stress $\sigma_{xy}$ [panel (a)] and osmotic pressure $P$ 
  [panel (b)] with P{\'e}clet number $Pe$ in forward shear. Panel(b) also shows the 
  scaling of $G_\infty \mathcal{N}_{1,2}/\sigma_{xy}^2$ with $Pe$. The effective scaling 
  laws are found to be $\sigma_{xy}\sim(Pe)^{0.36}, N_1\sim(Pe)^{0.51}$, $N_2\sim(Pe)^{0.58}$ 
  and $P\sim(Pe)^{0.37}$ respectively. A total of $400$ independent simulation runs are 
  averaged for strain rates below $5\times10^{-4}$ for statistics while $100$ averages 
  were sufficient for higher strain rates to obtain the graphics.}
\end{figure}

Fig.(\ref{fig:scaling}) shows the flow curve, that is the steady state scaling of the shear 
stress, first and second normal stresses and the osmotic pressure with the corresponding 
P{\'e}clet number ($Pe$) in forward shear at fixed temperature $T=0.4$. The structural 
relaxation timescale of the supercooled melt at this temperature is computed to be 
$\tau_{\alpha}=1.8\times10^3$, which is slower by a factor of $10^4$ than a single 
particle relaxation time. Panel (a) depicts an increase in $\sigma_{xy}$ with 
increasing $Pe$ which is a typical mechanical response of the melt, with a crossover 
from Newtonian to sub-Newtonian scaling regime for $Pe>0.1$, that corresponds to strain 
rates higher than $\dot\gamma = 10^{-4}$. It is to notice that above $T_c$, there exists 
no finite yield stress and the low $Pe$ response is always Newtonian, as been expected 
in a dense liquid mixture. However, we find that both first and second normal stresses 
remain in the sub-Newtonian regime for $Pe > 1$. As noted in fig.(\ref{fig:scaling}) 
caption on the power law scaling of stresses with $Pe$, scaling exponents of normal 
stresses are higher than the shear stress exponent. This results to a slower transition 
from Newtonian to sub-Newtonian regime for normal stresses than the shear stress, thus 
explaining the depicted behaviour of the stresses at moderate values of $Pe$. We expect 
the low shear rate response of both stresses to be Newtonian. In Panel (b), we depict 
the osmotic pressure as a function of $Pe$. The pressure decreases with decreasing $Pe$ 
and saturates for $Pe < 0.1$. The low $Pe$ response of pressure is thus independent 
of $Pe$ and reaches to the quiescent state value. On the other hand for moderately 
high values of $Pe$, pressure and shear stress grows in a similar fashion, reflected 
also in the power law scaling exponents. Similar response in pressure is also reported 
in simulations of hard-sphere glass \cite{MaGrRaVar}.  

Another quantity of interest is the dimensionless number $G_\infty \mathcal{N}_{1,2}/\sigma_{xy}^2$ 
with $G_\infty$ denoting the low frequency plateau modulus. For moderately high shear 
rates, shear-thinning generalized Maxwell model predicts, $\sigma_{xy} = G_\infty \gamma$ 
and $\mathcal{N}_{1,2} = 2G_\infty \gamma^2$ where $\gamma$ is the strain. Therefore 
$G_\infty \mathcal{N}_{1,2}/\sigma_{xy}^2 = 2$ \cite{PapVoigt}. $Pe$-dependence of this 
quantity is displayed in panel (b). For an increasing $Pe$, the ratio progresses 
towards a value $\sim 2$. This result thus restricts the applicability of the model 
at the considered range of $Pe$ numbers.
\begin{figure}
\centerline{\includegraphics[width=1.1\linewidth]{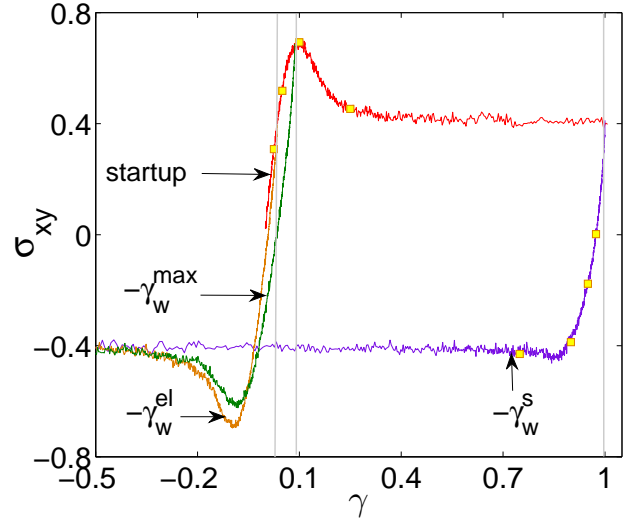}}
\caption{(Color online)\label{fig:stressstrain}
  Linear plot of stress-strain relation $\sigma_{xy}(\gamma)$ at fixed temperature 
  and strain rate for various flow histories: (i) starting from equilibrium (startup, 
  red), (ii) after flow reversal in the steady state (-$\gamma_w^s$, magenta), (iii) 
  from the elastic regime (-$\gamma_w^{el}$, brown) and (iv) from point of the stress 
  overshoot (-$\gamma_w^{max}$, green). Abscissa of the -$\gamma_w^s$ curve is shifted 
  from $\gamma=75$ to $\gamma=1$ for comparison. The vertical bars in gray denote the 
  successive points of shear-reversal from elastic, overshoot and steady state regime 
  of the startup curve. Square (yellow) dots, corresponding to the accumulated strain 
  indicated in fig.(\ref{fig:grtheta}), are printed on startup and steady state reversed 
  curve. A total of $200$ independent configurations are averaged to obtain the graphics.
}
\end{figure}

\subsubsection{Shear stress response}

Next we draw our attention in studying the transient response of the melt under forward 
and instantaneous reversal of the direction of applied shear. Fig.(\ref{fig:stressstrain}) 
depicts the shear stress as a function of strain $\gamma=\dot\gamma t$ for fixed strain 
rate $\dot\gamma$ and time $t$. Stress is measured after application of the strain 
rate to equilibrated configurations and after shear reversal from three different flowing 
states. In forward shear at a strain around $\gamma\approx10\%$, a profound stress overshoot 
from the steady state stress value is observed, found earlier in studies of binary colloidal 
melt with screened long ranged interaction \cite{zaholaegbrvofu}. State-of-the-art explanation 
of this behaviour follows as the enhancement of energies transferred by the shearing forces 
to the particles caged within their immediate neighbours with a characteristic length: the 
Lindemann length. At any strain lower than the denoted strain, the mechanical response of 
the mixture is that of an elastic solid with finite $G_\infty$ ($\sigma_{xy} \sim G_\infty\gamma$). 
Considering the time scales involved, the initial part of the $\sigma_{xy}(\gamma)$ curve 
for low strain rates is dominated by the long-time plateau modulus $G_\infty$ rather than 
the instantaneous modulus $G_0$. At large strains however, the response is that of a viscous 
fluid where $\sigma_{xy}$ is not a function of $\gamma$, but of $\dot\gamma$. Strains of 
order unity appear sufficient to drive the system into this state of steady flow. This 
is consistent with the expectation that the flow-induced decay of correlations occurs 
on a time scale set by $\dot\gamma^{-1}$. The argument leading to this state follows 
as the release of elastic energy from the breaking of local cages, that leads also to 
superdiffusive particle motion in between the ballistic and diffusive motion, reported 
in the study of mean squared particle displacement \cite{zaholaegbrvofu,frbhhofuvo}. 
Reversing the direction of the applied shear at various strains along the stress-strain 
curve results in an accumulated strain that first decreases to zero. After that, $|\gamma|$ 
grows linearly as a function of time. In steady flowing state, $|\gamma|\to\infty$ 
asymptote is $-\sigma_{\text{s}}(\dot\gamma)$ where $\sigma_{\text{s}}$ is the steady 
state stress at forward shear. Hence in fig.(\ref{fig:stressstrain}), different 
$\sigma_{xy}(\gamma)$ curves in shear reversal coincide at large strains. 

Comparing the steady state reversal curve with the initial startup curve, the most 
striking difference is the absence of a stress overshoot. This agrees with the findings 
of ref.\cite{kalepr1}, where the same phenomenology was reported for a system quenched 
initially into the glassy state. As a consequence, the steady-state value of the stress 
is reached earlier than the startup case at a value, slightly higher than the accumulated 
strain of $|\gamma|\approx 0.1$. This is somewhat surprising, since one might expect the 
pre-sheared state to bear structural anisotropies that accommodate flow in the $+x$ 
direction and hence oppose that in the $-x$ direction more than the isotropic equilibrium 
structure. This appears not the case in our simulation, that we elaborate more while 
discussing on the local microstructure. This is corroborated also by looking at the instantaneous 
effective elastic modulus, $G_\text{eff}=d\sigma/d\gamma|_{|\gamma-\gamma_0|=0.025}$ 
($\gamma_0$ being the strain when stress first decreases to zero): a lower value is 
found following flow reversal than the one characterizing the initial startup from 
the equilibrium configuration \cite{frbhhofuvo}. 

Flow reversal inside the elastic transient results in a stress-strain curve that exhibits 
a negative stress overshoot, displayed in fig.(\ref{fig:stressstrain}) labeled with 
$-\gamma_w^{el}$. The magnitude of the overshoot is identical to the initial startup flow. 
This exemplifies that during the elastic part, strain-induced rearrangements are essentially 
reversible, unlike in the plastic regime of viscous flow. The figure also depicts the case 
of flow reversal, once the initial stress-strain curve has reached its maximum (curve labeled 
with $-\gamma_w^{max}$). This intermediate case still exhibits a pronounced overshoot, albeit 
lower than the $-\gamma_w^{el}$ case. Thus, up to the stress overshoot, the response of the 
system to the initial flow is mainly reversible. This is consistent with the notion that the 
overshoot marks the nontrivial breaking of nearest neighbour cages due to the imposed flow. 
We want to remind the reader that an overshoot in $\sigma_{xy}(\gamma)$ implies a dynamic 
shear modulus or microscopic stress autocorrelation function that exhibits overrelaxation: 
stresses do not simply decorrelate but, during breaking of cages, are released in such a 
way that they briefly become anti-correlated during the process \cite{zaholaegbrvofu}.
\begin{figure}
\centerline{\includegraphics[width=1.0\linewidth]{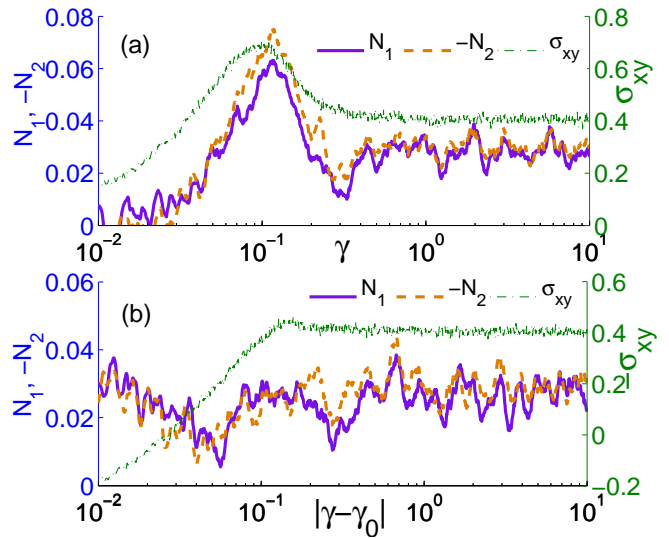}}
\caption{(Color online)\label{fig:normalstress}
  Semilog plot of the normal stress-strain relation $\mathcal{N}_{1,2}(\gamma)$ at fixed 
  temperature and strain rate for two different flow histories: panel (a): shear start up 
  from a quiescent state and panel (b): shear reversal from a steady flowing state (denoted 
  with -$\gamma_w^s$ in fig.\ref{fig:stressstrain}). For comparison, shear stresses in both situations are 
  depicted. A total of $1000$ independent simulation runs are averaged to obtain the graphics.}
\end{figure}

\subsubsection{Transient dynamics of normal stresses}

To complete the discussion on stresses, we study the transient and steady state response of 
normal stresses, defined in Eq.(\ref{eq.nstress}), as a function of strain. Panel (a) of 
fig.(\ref{fig:normalstress}) depicts $\mathcal{N}_1$ and $\mathcal{N}_2$ as a function of 
strain $\gamma$ for the startup shear from quiescent configuration, while panel (b) display 
results in shear reversal from the steady state, after shear stress becomes negative. The 
magnitude of the normal stress is noted approximately one order smaller than that of 
shear stress, which can be anticipated to the quadratic scaling of normal stress with 
$Pe$. For startup case, $\mathcal{N}_1$ builds up to attain steady state value after 
exhibiting an overshoot around $10\%$ of strain, very similar to the shear stress-strain 
response. The amount of overshoot in shear or normal stresses stay constant compared to 
the steady state value. $\mathcal{N}_2$ however exhibits a negative stress overshoot with 
a crossover from transient to steady state. Surprisingly we find, $\mathcal{N}_1 \sim -\mathcal{N}_2$ 
and $\mathcal{N}_3=\mathcal{N}_1+\mathcal{N}_2 \sim 0$. After reversing the flow direction 
from steady flowing state, we find unchanged response of the normal stresses to shear. 
Nonzero values of $\mathcal{N}_2$ indicates that in forward shear, the stress overshoot 
phenomena is not limited only to the shear direction, but also has a signature in two 
perpendicular directions to it. Unchanged magnitude and sign of normal stresses at shear 
reversal from steady state are indicative of uniform normal forces, devoid of the flowing 
direction. The distinction of history dependent response in flow reversal is not primarily 
due to the swapped flow direction, but due to whether the microscopic structure is still 
close to its equilibrium configuration (and hence only deformed reversibly), or whether 
it is sufficiently close to the flowing configuration.
\begin{figure}
\centerline{\includegraphics[width=1.0\linewidth]{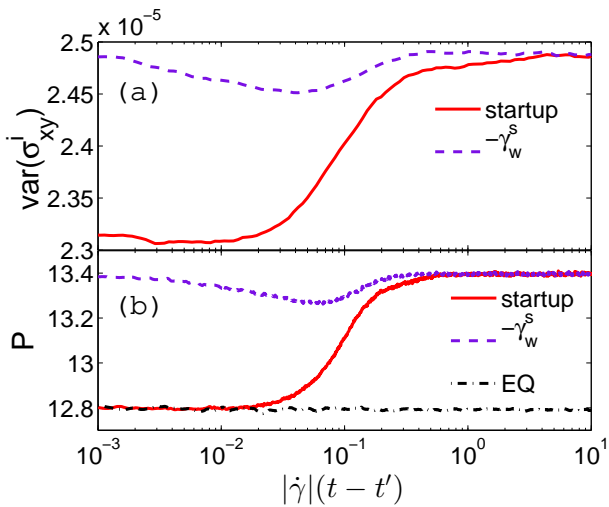}}
\caption{\label{fig:locstress}(Color online) Panel (a): Variance of the local stress distribution 
  corresponding to the stress-strain curves shown in fig.(\ref{fig:stressstrain}). Panel (b): 
  Osmotic pressure for the shear startup and shear reversal from the steady state. Pressure 
  at quiescent state is also marked for comparison. A total of $1000$ independent configurations 
  are averaged to obtain the graphics. 
}
\end{figure}

\subsubsection{Local stress fluctuation and pressure}

Further support for the hypothesis claimed in the above section comes from a study of the 
local stress fluctuations, initially suggested by Zausch {\it et al} \cite{zaho}. Defining 
a local stress element as $\sigma_{xy}^i=-(1/V)\sum_{j\neq i}r_{ij,x}F_{ij,y}$ such that 
$\langle\sigma\rangle=\sigma_\text{pot}$ is the potential part of the macroscopic stress, 
the distribution of these local stresses around their average value is computed. The upper 
panel (a) in fig.(\ref{fig:locstress}) depicts the variance of $\sigma_{xy}^i$ as a function 
of strain $|\gamma|$ in forward shear from quiescent state and instantaneous reversal from 
steady flowing state. As already noted in ref.~\cite{zaho}, the initial equilibrium 
configuration is characterized by a variance that is significantly lower than that in 
the flowing steady state. A steep increase in stress variance around $\gamma\approx0.1$ 
is found that coincides with the stress overshoot, separating the reversible elastic 
regime from the irreversible plastic counterpart. Reversing the shear flow from the 
steady state regime (denoted with $-\gamma_w^{s}$), $\var(\sigma_{xy}^i)$ essentially 
remains at the previously reached level, after exhibiting a small dip below $\gamma\leq0.1$. 
In fig.\ref{fig:locstress}(b), the osmotic pressure is displayed for the flow startup 
and reversal from steady state. For the startup case, the increase in the osmotic pressure 
from one state to another is found without any overshoot at strain $\approx 0.1$. In case 
of shear reversal, the value remains constant at the previously attained value with a small 
dip. A plausible explanation for the dip can be sought by an argument that instantaneous 
shear reversal interrupts the established planar Couette flow, thus eventually lowering 
the fluctuations. The stress variance as well as pressure reaches the forward steady 
state value as soon as the linear flow profile is re-established in the opposite 
direction. Identical response of pressure and the fluctuation in particulate stress 
with strain indicate of an interconnection between these seemingly different quantities. 

\subsection{Microstructure}

\subsubsection{Anisotropies in the transient structure}

\begin{figure}
  \subfigure[$\gamma=0.025$]{
  \includegraphics[width=1.2in, angle=0]{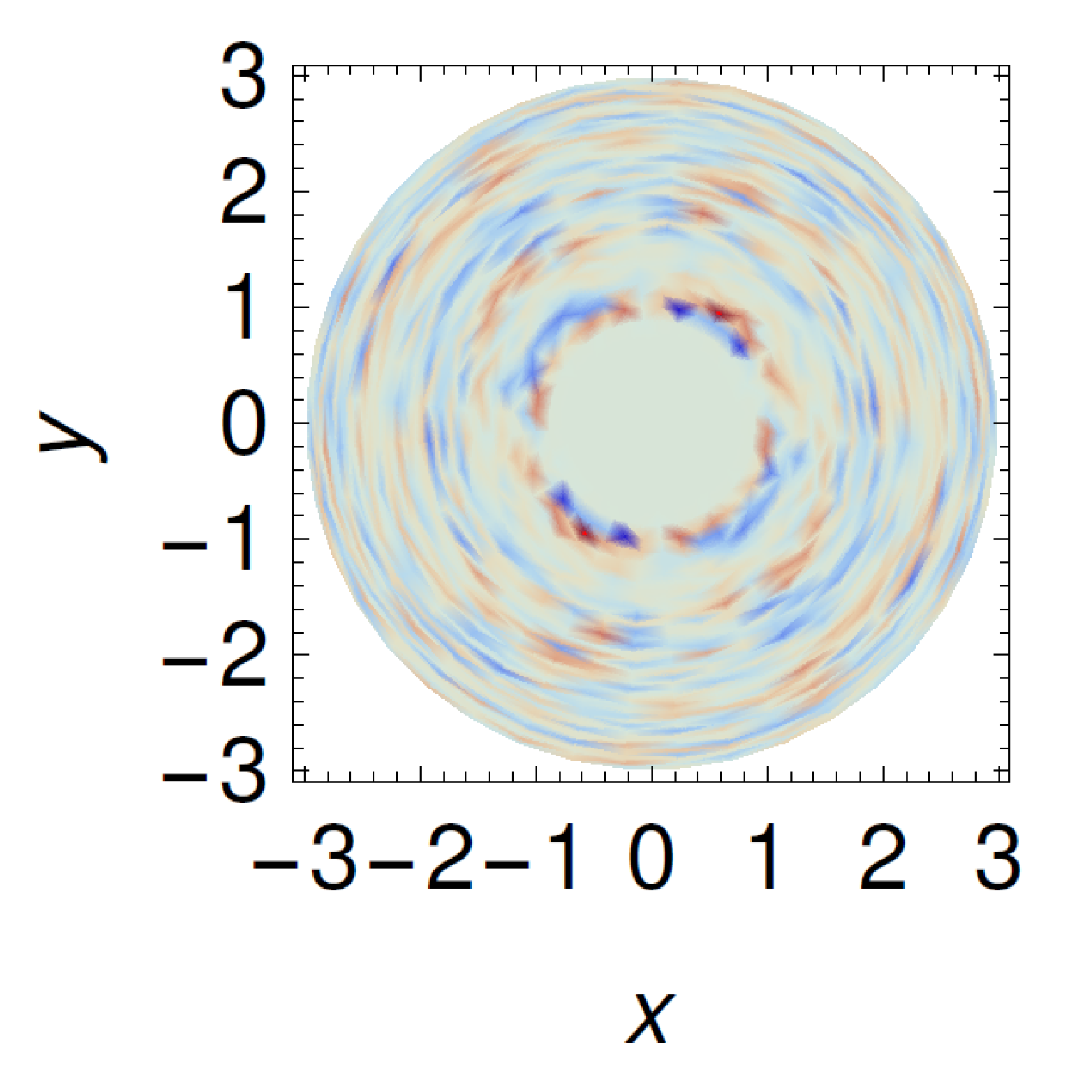} 
  }
  \subfigure[$\gamma=0.05$]{
  \includegraphics[width=1.2in, angle=0]{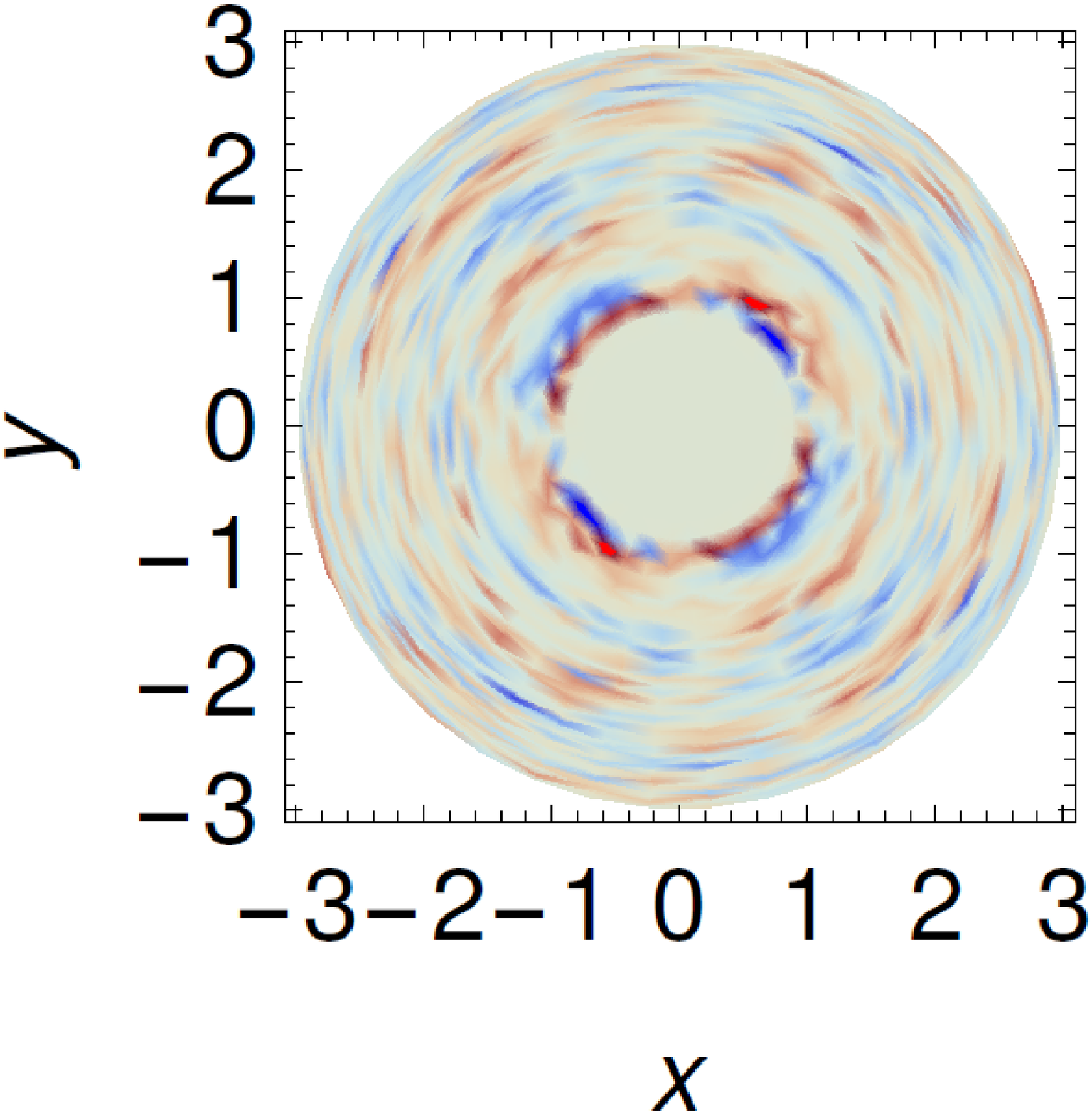} 
  }
  \subfigure[$\gamma=0.1$]{
  \includegraphics[width=1.2in, angle=0]{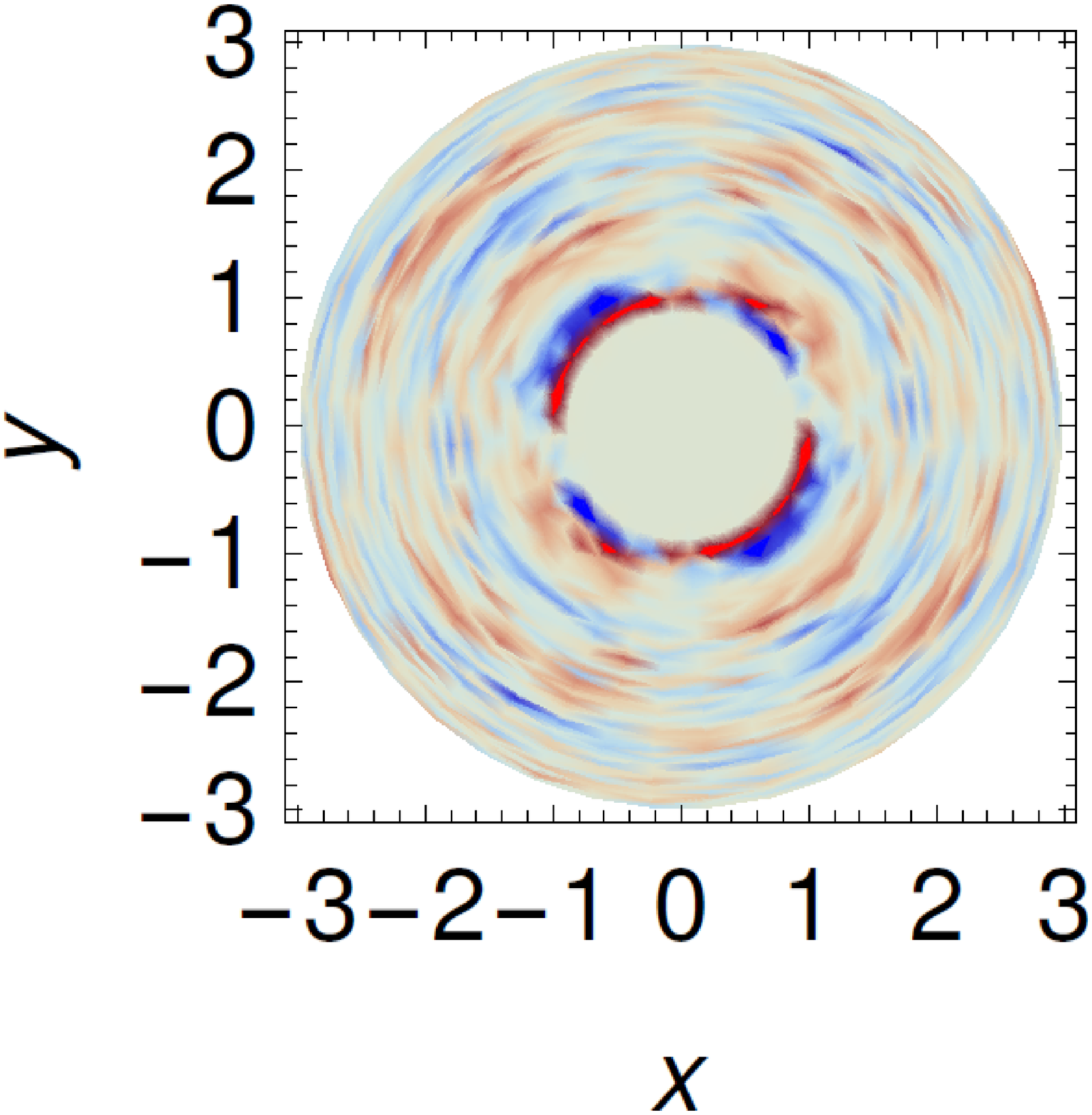} 
  }
  \subfigure[$\gamma=0.25$]{
  \includegraphics[width=1.2in, angle=0]{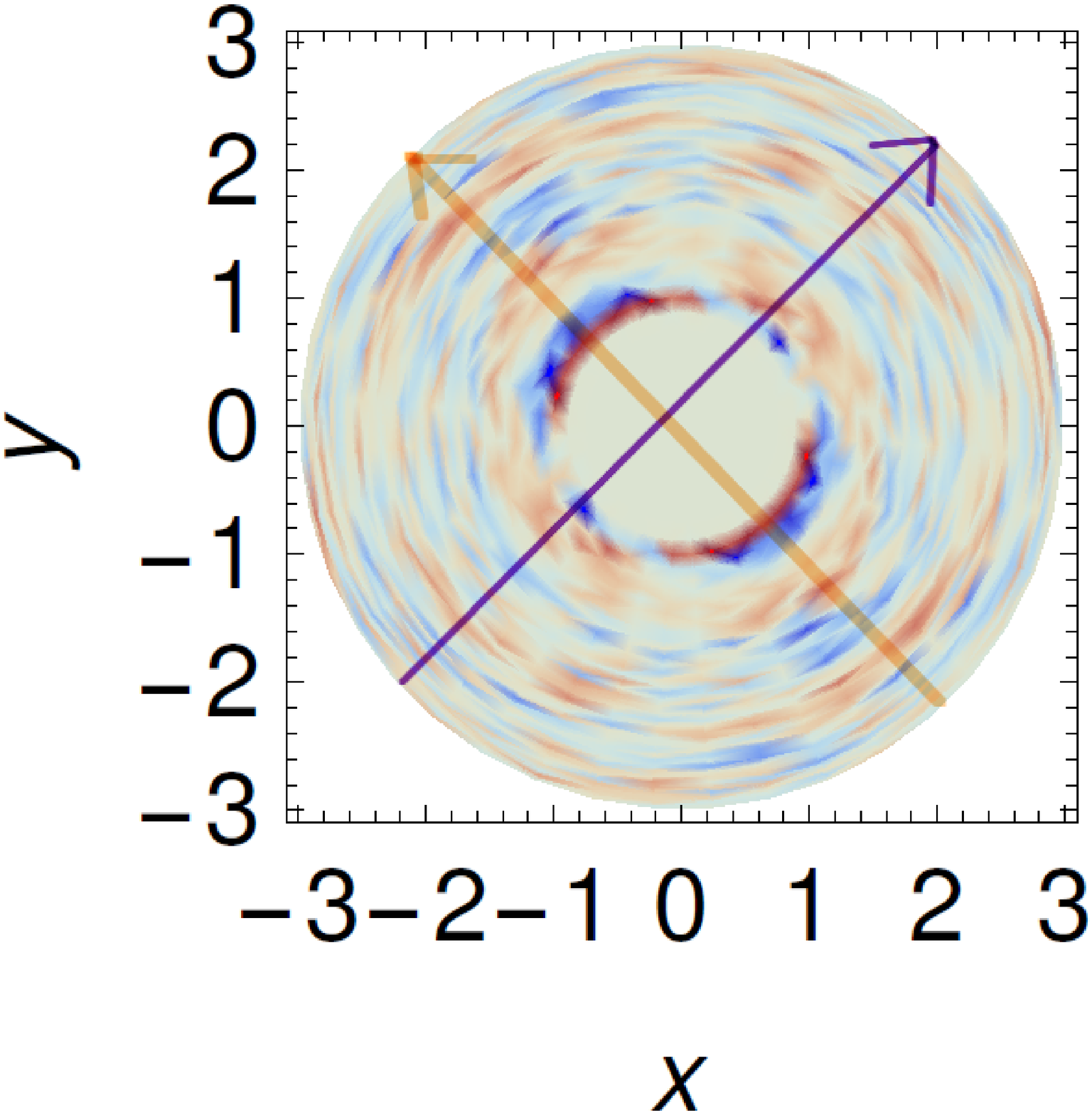}
  }

  \subfigure[$\gamma=74.975$]{
  \includegraphics[width=1.2in, angle=0]{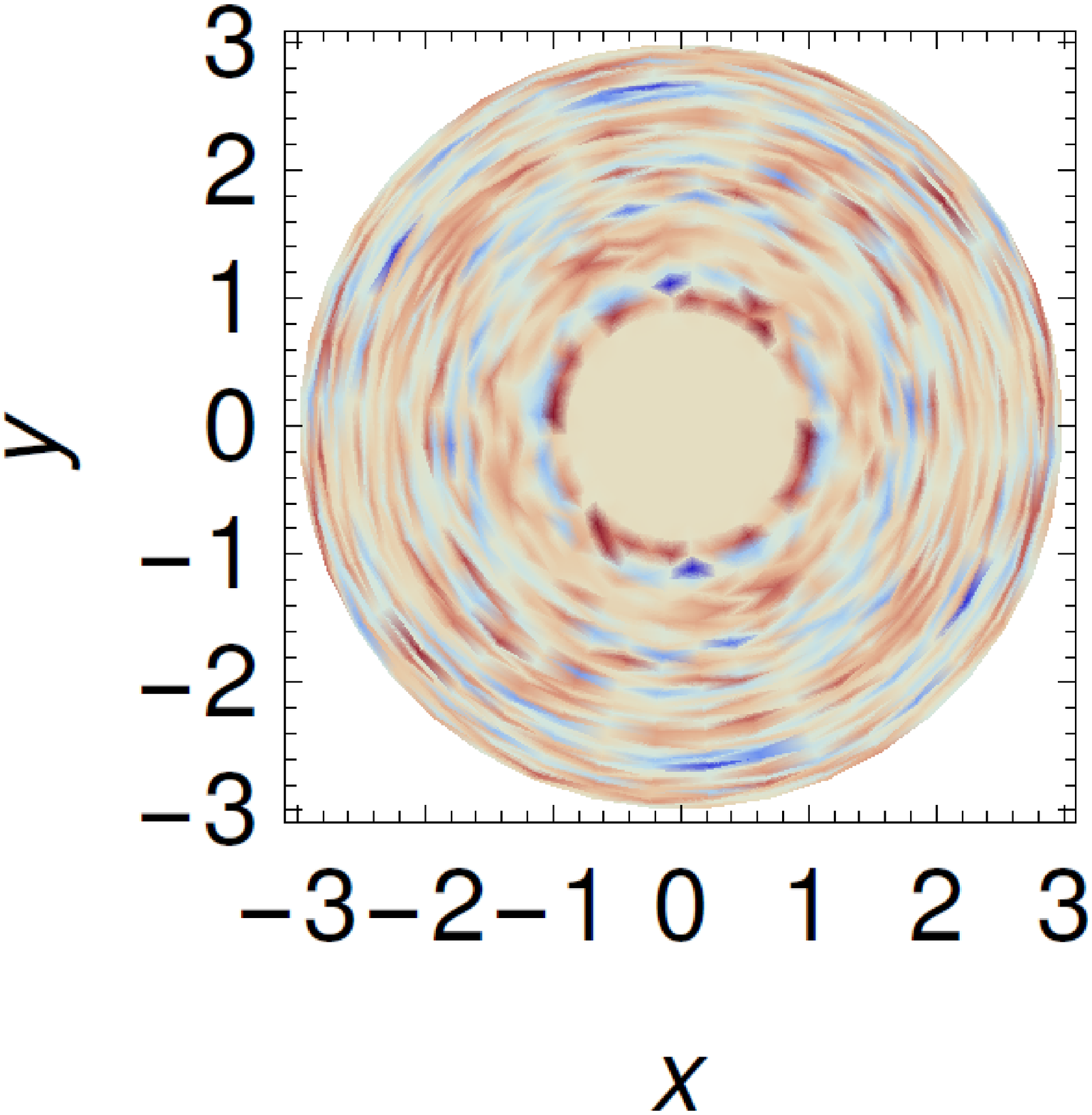} 
  }
  \subfigure[$\gamma=74.95$]{
  \includegraphics[width=1.2in, angle=0]{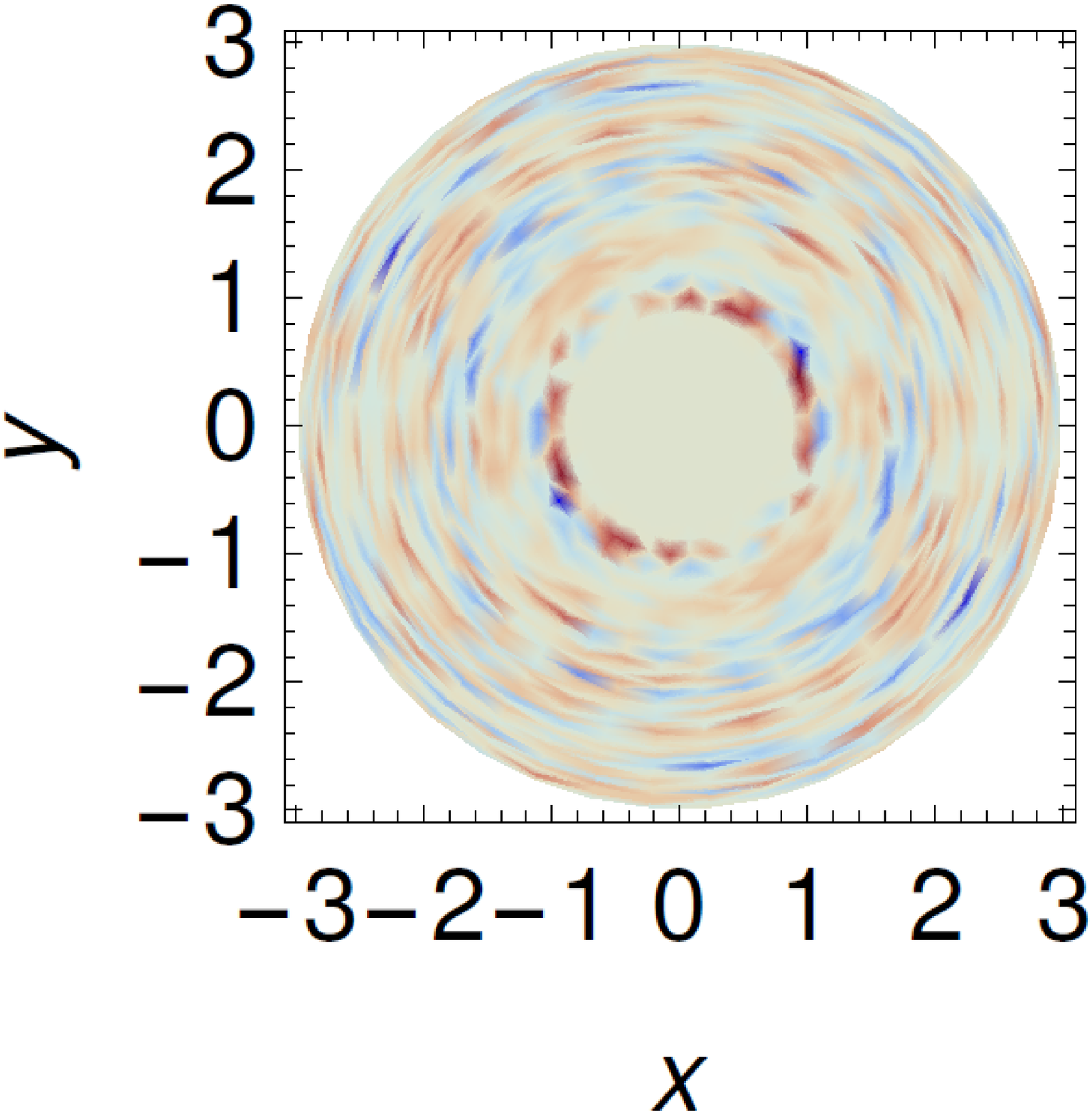} 
  }
  \subfigure[$\gamma=74.9$]{
  \includegraphics[width=1.2in, angle=0]{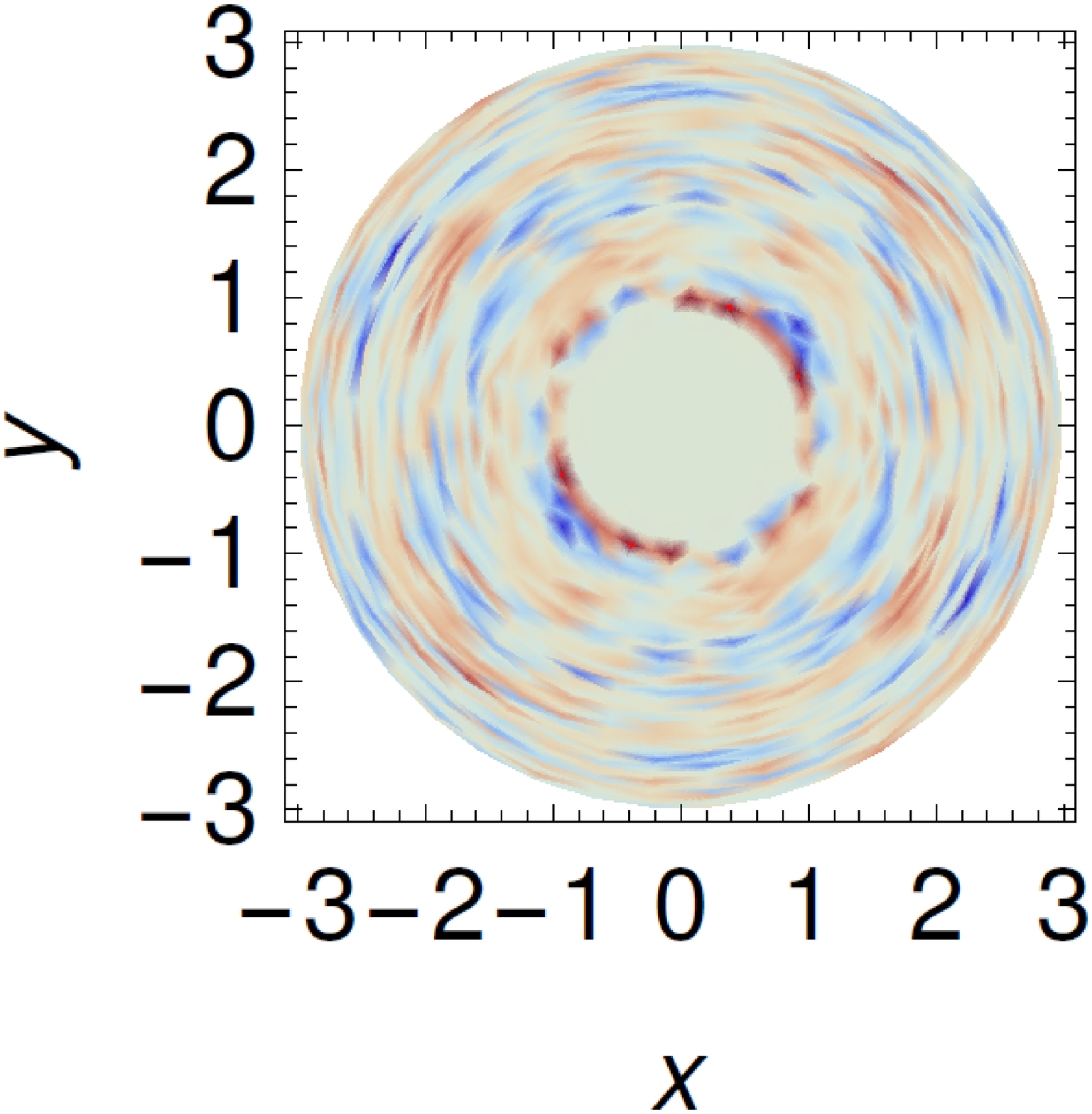} 
  }
  \subfigure[$\gamma=74.75$]{
  \includegraphics[width=1.2in, angle=0]{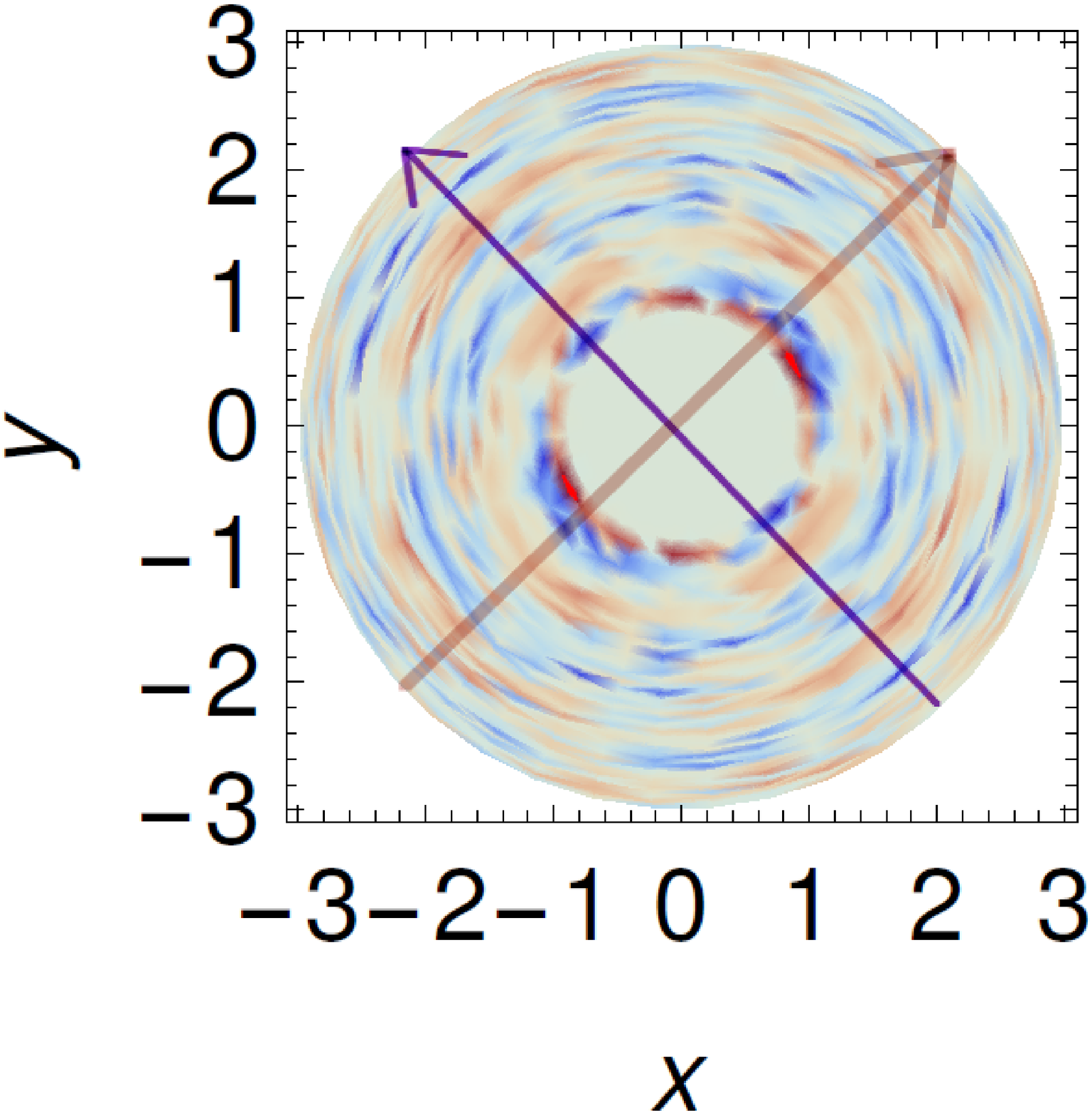}
  }
  
  \subfigure{
  \includegraphics[width=2.0in, height=0.3in, angle=0]{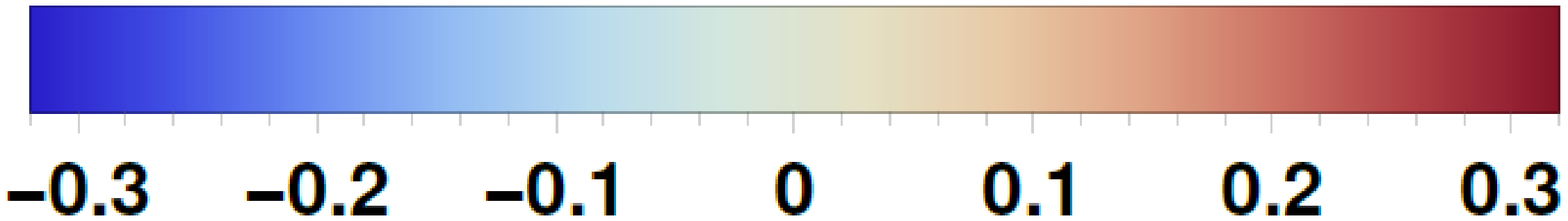} 
  }
\caption{\label{fig:grtheta}(Color online) Density plot of $g_{shear}(r,\theta)-g_{EQ}$ 
    in the shear-gradient plain depicted in false colour. States after shear startup 
    from a quiescent state with accumulated strains are indicated in panel (a-d) while 
    panel (e-h) show different states after shear reversal from plastic state with 
    accumulated strain $\gamma=75$. In panel (d) and (h), extension axis is displayed with 
    faint, brown arrow while compression axis is displayed with bold, magenta arrow. A total 
    of $100$ independent simulation runs are averaged and within each run $10$ configurations 
    centered around the respective strain value are averaged to obtain the graphics.}
\end{figure}

To understand the shear induced anisotropy in the local fluid structure, we measure the 
angle dependent radial distribution function $g(r,\theta)$ on shear-gradient (x-y) plain. 
Fig.(\ref{fig:grtheta}) shows the extra contribution of shear to the microstructure from the 
homogeneous and isotropic quiescent state, denoted with $g_{EQ}$. A faint, brown arrow 
printed on panel (d) and (h) shows the extensional axis while a bold, magenta arrow 
perpendicular to it in the same panel depicts the compressional axis. Successive points 
corresponding to the accumulated strain mentioned in fig.(\ref{fig:grtheta}), is printed in 
the stress-strain curve (fig.\ref{fig:stressstrain}) for a direct comparison. It is worth 
mentioning that planar Couette flow is established for all of the reported accumulated 
strains. As depicted in panel (a-d), in forward shear, the homogeneous state is steadily 
deformed with development of anisotropies along the extension-compression axis. This 
feature of the shear induced structural deformation that pushes more particles around 
the compressional axis while pulling the particles apart along the extensional axis is 
also reported in molecular simulations \cite{HanRaHess,koulauEgbraPet}. As could be 
anticipated from the stress-strain relationship, the amount of compression-extension 
is the maximum at the point of stress overshoot (shown in panel (c)). Panel (e-h) depict 
the time evolution of the structure at reversed flowing state. Instantaneous shear reversal 
from steady state results to an exchange between extension-compression axis with an 
intermediate uniformly compressed state, seen in panel (e). This clearly indicate of 
an absence of force chains or jamming of colloidal particles \cite{cawiboucla} in the 
considered shear rate. The vanishing of the stress overshoot attributes to ceasing of 
maximal anisotropy along the compression-extension axis present in forward shear. Finally, 
panel(d) and panel (h) confirms the equivalence of steady state structure in opposite 
flowing directions without any memory of the flow history. 
 
\begin{figure}
\centering
\includegraphics[width=1.0\linewidth]{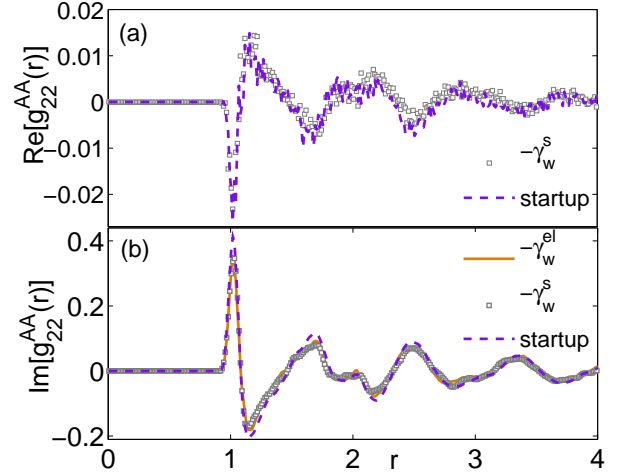}
\caption{(Color online) Panel(a) shows steady state $Re[g_{22}^{\alpha\alpha}(r)]$ 
    correlations between $\alpha \in {A}$-particles, obtained by correlating $250$ 
    steady state configurations within a strain window $[50,75]$ in a single run and 
    then averaging over $20$ independent samples for shear start up (magenta, dashed 
    line) and reversal from plastic steady state (square symbol). Panel(b) shows the 
    imaginary part of $g_{22}^{\alpha\alpha}(r)$ correlations between $\alpha \in 
    {A}$-particles after shear reversal from elastic (brown, solid line) and plastic 
    branch (square symbol) at steady state shear stress $\sigma_{xy}=-0.36$. Sign 
    reversed forward steady state curve (magenta, dashed line) is also shown for 
    comparison.}
\label{fig:g22AA}
\end{figure}

\subsubsection{Steady state structure}

To conclude the discussion on microstructure, we calculate the steady state structure 
observed in the local projected quantities $Re[g_{22}^{\alpha\alpha}(r)]$ and 
$Im[g_{22}^{\alpha\alpha}(r)]$ defined in Eq.(\ref{eq:Reg22}-\ref{eq:Img22}). Recall 
that the first normal stress can be obtained from $Re[g_{22}^{\alpha\alpha}(r)]$ as 
$\mathcal{N}_{1}=\rho^2\sqrt{{2\pi}/{15}}\sum_{\alpha\beta}{N_\alpha N_\beta}/{N^2} 
\int_0^\infty dr r^3 [{\partial V_{\alpha\beta}(r)}/{\partial r}] \times Re[g_{22}^{\alpha\beta}(r)]$.
The upper panel of fig.(\ref{fig:g22AA}) compares the steady state structure 
$Re[g_{22}^{\alpha\alpha}(r)]$ for the shear startup from a quiescent state 
as well as the shear reversal from the steady state for $\alpha \in {A}$-species. 
We find excellent agreement between the structure in steady state for both directions, 
indicative of a homogeneously flowing state with equal planar anisotropy devoid of 
any memory of the quiescent or pre-sheared configuration. The most important contribution 
to first normal stress is attributed to the nearest neighbour distance structure of the 
pair correlator. As seen for nearest neighbour distances, particles are squeezed along 
the compressional axis in the plain and extended along the extensional axis. This results 
to a positive $\mathcal{N}_1$ at the considered strain rate. 

The shear stress on the other hand can be derived from $Im[g_{22}^{\alpha\alpha}(r)]$ as
$\sigma_{xy} = -\rho^2\sqrt{{2\pi}/{15}}\sum_{\alpha\beta}{N_\alpha N_\beta}/{N^2} 
\int_0^\infty dr r^3 [{\partial V_{\alpha\beta}(r)}/{\partial r}] \times Im[g_{22}^{\alpha\beta}(r)]$.
Panel (b) compares the positional dependence of $Im[g_{22}^{\alpha\alpha}(r)]$ for 
$\alpha \in {A}$-species at steady state for two different shear reversal state, denoted 
by $-\gamma_w^{el}$ and $-\gamma_w^{s}$, to the forward directed steady state. The 
correlator (sign reversed) in forward shear agrees well with the correlator obtained 
from the steady reversed flowing states. This result coincides with an earlier claim 
that at equal moduli of shear stress, the projected structure retains it's shape devoid 
of the flow history \cite{zaho}. We found this to hold for reverse flowing states as 
well and thus extending the claim for first normal stress-structure relation at steady 
state and numerically validate Eq.(\ref{eq:N1}).

\section{Conclusions and Outlook}
\label{sec:5}

By employing a dissipative particle dynamics scheme in conjunction to Lees-Edwards 
boundary condition to soft repulsive colloids, in this article we have discussed the 
nonlinear rheology of dense colloidal melt under shear flow, specifically the transient 
and steady state properties after a sudden application of steady strain rate $\dot\gamma$, 
starting from either a quiescent state or various configurations that have been pre-sheared 
in the opposite direction. Functional dependence of shear and normal stresses as well as 
osmotic pressure with P{\'e}clet number is sought. A crossover from Newtonian to sub-Newtonian 
regime is found in shear-stress for $Pe>0.1$ while the normal stresses remain in the 
sub-Newtonian regime. The osmotic pressure saturates for lower $Pe$. The binary melt 
exhibit shear thinning and for much higher strain rates can result into a shear thickening 
behaviour with a negative $\mathcal{N}_1$ \cite{mewagner}. 

Stress-strain curves of the pre-sheared configurations are measured along with shear 
start-up. In addition to the overshoot in shear stress, overshoot in first and second 
normal stresses is observed at 10\% strain amplitude with a step jump in the osmotic 
pressure as well as in the particulate stress variances. However, once attained the 
steady state, no stress overshoot is found while unchanged state of pressure and 
local stress variance emerge in response to shear reversal. This validates the conjecture 
of shear induced nearest-neighbour cage breaking at startup flow, having a steady 
state with weak cages that ceases to play any dramatic role when the direction of 
flow is reversed. An interesting connection can be sought between the fluctuations 
of the particulate shear and normal stresses with the osmotic pressure, which is 
beyond the work presented here.
 
Angle dependent pair correlation function depicts that a uniform exchange of 
compression-extension axis occurs with a continuous evolution of structure, obtaining 
a steady Couette flow in relatively short span of time without cluttering or formation 
of force chains which are a typical signature in athermal systems. We also do not find 
any shear induced crystallization at the considered shear rates. Maximal anisotropy is 
exhibited at a strain where stress overshoot appears in forward shear while such maxima 
cease to exist in the shear reversed states. The steady state structure in both direction
containing equal anisotropy confirms an absence of flow history in steady state. The 
agreement in shear induced structure at steady flowing state validates the relationship 
between various components of the pair correlator and macroscopic stresses. The change 
of sign in shear direction attributes to an exchange of extension-compression axis as 
noted in the imaginary component of the pair distribution function while unchanged 
structure of the real component of the pair correlator confirms of a positive $\mathcal{N}_1$ 
for both directions of shear. In a similar spirit to Eq.(\ref{eq:N1}-\ref{eq:sigma}), 
relation between $\mathcal{N}_2$ and $\mathcal{P}$ to the components of pair correlator 
can be obtained \cite{ganeu} and an unchanged structure at equal stresses can be found. 
However, such studies should be performed at larger strain rates with the melt been 
quenched deep into the glassy state, which is outside the scope of the presented work. 
We seek for experimental measurement in dense supercooled melt to verify the claims 
presented in this article.
 
\begin{acknowledgments}

We thank Th. Voigtmann for several insightful suggestions and M. Fuchs, N. Wagner and R. Adhikari 
for useful discussions. We also thank P Kuhn for providing routine to calculate the angle dependent 
radial distribution function. We gratefully acknowledge funding through the Helmholtz Gesellschaft 
(HGF, VH-NG~406). 
\end{acknowledgments}

\bibliographystyle{apsrev}
\bibliography{references}
\end{document}